\documentstyle[mprocl]{article}

\def\be{\begin{equation}}
\def\ee{\end{equation}}
\def\bea{\begin{eqnarray}}
\def\eea{\end{eqnarray}}
\def\lb{\label}
\begin{document}

\title{REGULARITY AND STABILITY OF ELECTROSTATIC SOLUTIONS IN
KALUZA-KLEIN THEORY}

\author{ M. AZREG-A\"{I}NOU}

\address{Girne American University,
Faculty of Engineering, Karmi Campus Karaoglanoglu,
Girne, North Cyprus (via Mersin 10, Turkey)\\E-mail:
azreg@taloa.unice.fr} 

\author{ G. CL\'EMENT}

\address{  
LAPTH (CNRS), B.P.110, 
F-74941 Annecy--le--Vieux cedex, France\\E-mail: gclement@lapp.in2p3.fr}  

\author{C.P. CONSTANTINIDIS, J.C. FABRIS}
\address{Dpto de F\'{\i}sica, Universidade Federal do 
Esp\'{\i}rito Santo, Vit\'oria, ES, Brazil\\E-mail:
clisthen@cce.ufes.br,   fabris@cce.ufes.br}

\maketitle\abstracts{
We investigate the family of electrostatic spherically symmetric solutions 
of the five-dimensional Kaluza-Klein theory. Besides black holes and 
wormholes, a new class of geodesically  complete solutions is
identified. A monopole perturbation is carried out, enabling us to
prove analytically the stability of large classes of solutions,
including all black holes and neutral solutions.}

Five--dimensional Kaluza--Klein theory, or sourceless general
relativity in 4+1 spacetime dimensions is historically one of the
first unified field theories, and deserves to be investigated as
the prototype of other multidimensional theories. The static, spherically
symmetric solutions of Kaluza--Klein theory have been obtained
independently by several authors \cite{Leut,DM,CD}. These solutions
include regular black holes, which were investigated by Gibbons and Wiltshire
\cite{GW}. A class of regular, horizonless charged solutions with wormhole
spatial topology was also identified by Chodos and Detweiler \cite{CD}.
The aim of the present work is to analyze more fully the geometrical
properties of the static spherically symmetric solutions of Kaluza--Klein
theory, as well as to investigate their stability \cite{kk4}.

Following the approach of Dobiasch and Maison \cite{DM}, the
spherically symmetric, electrostatic solutions of Kaluza--Klein theory
are given by
\be\lb{sol}
ds^{\,2} = -{|\lambda(r)|}^{-1}(dr^2 + H(r)d\Omega^2) + 
\lambda_{ab}(r)dx^adx^b ,
\ee
($a,b = 4,5$, with $x^4 = t$), where
\be
\lambda = \eta e^{N\sigma}, \quad 
\sigma = \frac{1}{2\,\nu}\,\ln\left(\frac{r-\nu}{r+\nu}\right), \quad
H = r^{2} - {\nu}^{2},
\ee
and the matrices 
\be
\eta =  \left( \begin{array}{cc}
                     1 & 0 \\
                     0 & -1 \\
		 \end{array} \right), \quad
N = \left( \begin{array}{cc}
            x-a & b \\
            -b & a 
    \end{array} \right)
\ee
depend on three real parameters $x$, $a$ and $b$ (proportional to the
scalar and electric charges), and the auxiliary real parameter
${\nu}^{2} = (x^{2} - y)/4$, where $y = b^2 + ax - a^2$.

These solutions correspond to black holes \cite{GW} with event horizon 
$r = \nu$ for $y = 0$, $x > 0$, $a \le 0$. Another class of regular
solutions, traversable wormholes \cite{CD}, occur for $y > x^2$. 
Extrapolation from the diagonal, uncharged case seems to lead to
the conclusion \cite{GW} that all the other cases $y \le x^2, y \neq
0$ correspond to naked singularities. However for $y \ge x^2/4$, the
matrix $\lambda$ cannot be diagonalized, so that the extrapolation is
invalid. We have found that the 5--dimensional Kretschmann scalar
behaves near the Killing horizon $r = \nu$ as $K \propto (x-2\nu)^2 
(r-\nu)^{(x-4\nu)/\nu}$.
This is finite for $x = 2\nu$ ($y = 0, x > 0$), corresponding to black
holes or conical singularities, as well as for $\nu \le x/4$ ($3x^2/4 \le y <
x^2$). This observation is reinforced by the analysis of geodesic motion in 
the 5-dimensional geometry (\ref{sol}), which leads to the conclusion
that for $y \ge 3x^2/4$ the Killing horizon is at infinite affine 
distance, so that these spacetimes are geodesically complete. 

To address the problem of linearization stability under monopole
perturbations, we linearize a time-dependent spherically symmetric solution
\be\lb{timedep}
ds_5^2 = e^{2\gamma(\rho,t)}\,dt^2 -  e^{2\alpha(\rho,t)}\,d\rho^2 -
 e^{2\beta(\rho,t)}\,d\Omega^2 -  e^{2\psi(\rho,t)}(dx^5 +
2V(\rho,t)\,dt)^2
\ee
around the static solution (\ref{sol}) as $\psi (\rho,t) = \psi_0(\rho) + 
\delta\psi(\rho)e^{kt}, ...$, and separate the linearized
5-dimensional Einstein equations in an appropriate gauge. In the
simple case of electrically neutral solutions ($b = 0$), the
linearized equation for the perturbation
$\delta\psi$ in the harmonic gauge $\delta\alpha - 2\delta\beta - 
\delta\gamma - \delta\psi= 0$ can be written as a Schr\"{o}dinger-like 
equation 
\be\lb{mast}
\delta\psi''(\rho) - U(\rho)\delta\psi(\rho) = 0
\ee
with positive-definite potential and zero energy, leading us to
the conclusion that these solutions are stable. The discrepancy with
the previous analysis \cite{Tom} of Tomimatsu (who found that the only stable
neutral solution was the 5--dimensional embedding of the Schwarzschild
black hole) is due to his use of weaker boundary conditions. 

In the case of generic charged solutions ($b \neq 0$), the linearized
equations are separated in the gauge $2\delta\beta + \delta\psi = 0$,
leading to a Schr\"{o}dinger-like equation (\ref{mast}) for
$\delta\psi$ with a non positive-definite effective potential
$U(\rho)$. However we have found that in the two parameter domains
($y = 0,\,a < 0,\,x \ge 3a/4$) and ($0 < y \le x^2/4,\,-2|b| \le a <
-|b|$) the potential function is positive definite, leading to the
stability of the corresponding static solutions. The first stability
class includes among others all the black hole and extreme black
hole solutions. The remaining cases (including the
geodesically complete solutions) lead to eigenvalue problems, which
should be solved numerically.

\eject
\end{document}